\documentclass[conference,letterpaper]{IEEEtran}
\usepackage{cite}
\usepackage{graphicx,color,epsfig,rotating}
\usepackage{amsfonts,amsmath,amssymb,bbm}
\usepackage{algorithm}
\usepackage{algpseudocode}
\usepackage{subfigure}
\usepackage{amsmath}
\usepackage{cite}
\usepackage{placeins}
\usepackage{graphicx}
\usepackage[latin1]{inputenc}
\usepackage{amssymb}
\usepackage{multirow}
\usepackage{stfloats}
\usepackage{tabularx} 
\usepackage{booktabs} 
\usepackage{url}
\usepackage{bm}
\usepackage{soul}
\usepackage{float}
\usepackage{lipsum}     
\usepackage{cuted} 
\usepackage{pstricks}
\usepackage{arydshln}
\usepackage{xspace} 
\usepackage[left=1.59cm,right=1.59cm, top=.77in, bottom=1.05in]{geometry} 

\allowdisplaybreaks  % global equation spacing
\newcommand{\user}[1]{User~$#1$}

\long\def\comment#1{}

\makeatletter

\newcommand{\thickhline}{%
    \noalign {\ifnum 0=`}\fi \hrule height 1pt
    \futurelet \reserved@a \@xhline
}
\newcolumntype{"}{@{\hskip\tabcolsep\vrule width 1pt\hskip\tabcolsep}}
\makeatother

\makeatletter
\newcommand{\subalign}[1]{
  \vcenter{%
    \Let@ \restore@math@cr \default@tag
    \baselineskip\fontdimen10 \scriptfont\tw@
    \advance\baselineskip\fontdimen12 \scriptfont\tw@
    \lineskip\thr@@\fontdimen8 \scriptfont\thr@@
    \lineskiplimit\lineskip
    \ialign{\hfil$\m@th\scriptstyle##$&$\m@th\scriptstyle{}##$\crcr
      #1\crcr
    }%
  }
}
\makeatother

\newtheorem{example}{Example}%[section]
\newtheorem{theorem}{Theorem}%[section]
\newtheorem{lemma}{Lemma}%[section]
\newtheorem{corollary}{Corollary}%[section]
\def \inputsum{{\sum_{i=1}^KW_i}}
\def \rzsigma{{R_{Z_{\Sigma}}}}

\def \rx{{R_X}} 

\def \rz{{R_Z}} 
\def \lx{{L_X}} 
\def \lz{{L_Z}}

\def  \zsigma{{Z_{\Sigma}}}
\def  \lzsigma{{L_{Z_{\Sigma}}}}

\let\tbf\textbf
\let\tit\textit

\let \bksl\backslash

\newcommand{\corr}{correlated\xspace}
\newcommand{\Iwf}{In what follows\xspace}

\newcommand{\Aar}{As a result\xspace}

\newcommand{\corrspdg}{corresponding\xspace}
\newcommand{\dsa}{decentralized secure aggregation\xspace}

\newcommand{\skr}{source key rate\xspace}

\newcommand{\eg}{e.g.\xspace}

\newcommand{\ie}{i.e.\xspace}
\newcommand{\msg}{message\xspace}
\newcommand{\msgs}{messages\xspace}
\newcommand{\Wlog}{Without loss of generality\xspace}

\newcommand{\hie}{hierarchical\xspace}
\newcommand{\secty}{security\xspace}

\newcommand{\Msp}{More specifically\xspace}

\newcommand{\Ip}{In particular\xspace}
\newcommand{\af}{as follows\xspace}
\newcommand{\resp}{respectively\xspace}
\newcommand{\iid}{i.i.d.\xspace}

\newcommand{\Thm}{Theorem\xspace}

\newcommand{\info}{information\xspace}

\newcommand{\itic}{information-theoretic\xspace}

\newcommand{\simuly}{simultaneously\xspace}
\newcommand{\etal}{\textit{et al.}\xspace}

\newcommand{\collu}{collusion\xspace}

\newcommand{\agg}{aggregation\xspace}

\newcommand{\secagg}{secure aggregation\xspace}

\newcommand{\Agg}{Aggregation\xspace}

\newcommand{\diff}{different\xspace}
\newcommand{\Diff}{Different\xspace}

\newcommand{\indep}{independent\xspace}

\newcommand{\indepce}{independence\xspace}

\newcommand{\indiv}{individual\xspace}
\newcommand{\Indiv}{Individual\xspace}
\newcommand{\comm}{communication\xspace}
\newcommand{\Comm}{Communication\xspace}

\newcommand{\achved}{achieved\xspace}
\newcommand{\achvb}{achievable\xspace}

\newcommand{\achvblty}{achievability\xspace}

\newcommand{\muinfo}{mutual information\xspace}

\newcommand{\Feon}{For ease of notation\xspace}

\newfont{\bbb}{msbm10 scaled 700}

\newfont{\bb}{msbm10 scaled 1100}

\newcommand{\kth}{{$k^{\rm th}$ }}

\newcommand{\Cc}{{\cal C}}

\newcommand{\Rc}{{\cal R}}
\newcommand{\Sc}{{\cal S}}
\newcommand{\Tc}{{\cal T}}

\newcommand{\eqdef}{\stackrel{\Delta}{=}}

\newcommand{\be}{\begin{equation}}
\newcommand{\ee}{\end{equation}}
\newcommand{\bea}{\begin{eqnarray}}
\newcommand{\eea}{\end{eqnarray}}

\begin{document}
\title{Information-Theoretic Secure Aggregation in Decentralized Networks}

\author{\IEEEauthorblockN{Xiang Zhang\IEEEauthorrefmark{1}, 
Zhou Li\IEEEauthorrefmark{2},
Shuangyang Li\IEEEauthorrefmark{1}, 
Kai Wan\IEEEauthorrefmark{3}, 
Derrick Wing Kwan Ng\IEEEauthorrefmark{4}, 
and Giuseppe Caire\IEEEauthorrefmark{1}
}
\IEEEauthorblockA{
Department of Electrical Engineering and Computer Science, Technical University of Berlin\IEEEauthorrefmark{1}\\
Guangxi Key Laboratory of Multimedia Communications and Network Technology, Guangxi University\IEEEauthorrefmark{2}\\
School of Electronic Information and Communications, Huazhong University of Science and Technology\IEEEauthorrefmark{3}\\
School of Electrical Engineering and Telecommunications, the University of New South Wales\IEEEauthorrefmark{4}\\
Email:~\IEEEauthorrefmark{1}\{xiang.zhang, shuangyang.li, caire\}@tu-berlin.de,~\IEEEauthorrefmark{2}lizhou@gxu.edu.cn,~\IEEEauthorrefmark{3}kai\_wan@hust.edu.cn,
\IEEEauthorrefmark{4}w.k.ng@unsw.edu.au}
}

\maketitle

\begin{abstract}
Motivated by the increasing demand for data security in decentralized federated learning (FL) and stochastic optimization, we formulate and investigate the problem of information-theoretic \emph{decentralized secure aggregation} (DSA). Specifically, we consider a network of $K$ interconnected users, each holding a private input, representing, for example, local model updates in FL, who  aim to simultaneously compute the sum of all inputs while satisfying the security requirement that
no user,  even when colluding with up to $T$ others,  learns anything beyond the intended sum. 
We characterize the optimal rate region, which specifies the minimum achievable communication and secret key rates for DSA. 
In particular, we show that to securely compute one bit of the desired input sum, each user must (i) transmit at least one bit to all other users, (ii) hold at least one bit of secret key, and (iii) all users must collectively hold no fewer than  $K - 1$ independent key bits.
Our result establishes the fundamental performance limits of DSA and offers insights into the design of provably secure and communication-efficient protocols for distributed learning systems.
\end{abstract}

\section{Introduction}
\label{sec:intro} 
Federated learning (FL) has emerged as a powerful paradigm for distributed machine learning (ML), enabling users to collaboratively train a global ML model without directly sharing their private datasets~\cite{mcmahan2017communication}. 
In the conventional centralized setting, a parameter server coordinates the model \agg and distribution process.
More recently, FL has been studied in a serverless 
setting known as decentralized FL (DFL)~\cite{lalitha2018fully,he2018cola, beltran2023decentralized,rodio2025whisper}, where model aggregation is distributed across neighboring nodes over a communication graph, thus enabling fast local updates. Compared with centralized FL, DFL offers several benefits, including improved fault tolerance and more evenly distributed communication and computation workload across the participating users~\cite{beltran2023decentralized}.

Although the local datasets are not explicitly shared, FL still exposes vulnerabilities in the presence of an untrusted server~\cite{bouacida2021vulnerabilities, geiping2020inverting, mothukuri2021survey}.
The demand for stronger security guarantees  has stimulated the study of \secagg (SA) in FL~\cite{bonawitz2017practical, bonawitz2016practical, 9834981,wei2020federated,hu2020personalized,zhao2020local,yemini2023robust,so2021turbo, liu2022efficient,jahani2023swiftagg+}.
In the pioneering work of Bonawitz \etal~\cite{bonawitz2017practical}, an SA protocol relying on pairwise generated secret keys to conceal the users' local models. 
Since then, various cryptographic methods based on random seed agreement have been employed to realize SA, typically under computational security. 

In a different vein, \emph{\itic} \secagg~\cite{9834981,zhao2023secure} investigates the fundamental limits of \secagg under \emph{perfect} security guarantee characterized by a zero mutual information criterion. \Ip, let the local models of FL be represented by  a set of input variables $W_1,\cdots, W_K$.
The \agg server aims to compute the sum of inputs $\sum_{k=1}^KW_k$ subject to the security constraint that it should not infer any \info about the inputs beyond their sum,  which can be characterized by  the zero mutual \info condition
$
I(\{X_k\}_{k=1}^K; \{W_k\}_{k=1}^K|\sum_{k=1}^KW_k)=0
$, where $X_k$ denotes the \msg upload from the \kth user. This ensures that  the received  messages by the server are statistically \indep of the inputs, thereby achieving perfect security. 

Recent studies on \itic SA have extended its scope to a wide range of practical constraints, including user dropout and collusion resilience~\cite{9834981, zhao2023secure,zhang2025secure,so2022lightsecagg,jahani2022swiftagg,jahani2023swiftagg+}, groupwise keys~\cite{zhao2023secure,wan2024information,wan2024capacity, zhang2025secure}, user selection~\cite{zhao2022mds, zhao2023optimal}, heterogeneous security~\cite{li2023weakly,li2025weakly,li2025collusionresilienthierarchicalsecureaggregation}, oblivious server~\cite{sun2023secure}, SA with multiple recovery goals~\cite{yuan2025vector}, and \hie secure \agg (HSA)~\cite{zhang2024optimal, 10806947,egger2024privateaggregationhierarchicalwireless, lu2024capacity,zhang2025fundamental, 11195652,li2025collusionresilienthierarchicalsecureaggregation}.
\Msp, \cite{9834981} studied SA in a server-based star network topology, accounting for both user dropout  and server-user collusion.   SA with improved key storage was studied in~\cite{ so2022lightsecagg,so2021turbo}.
SA with groupwise keys, where each user subset of a given size shares an independent key, was studied in~\cite{zhao2023secure,wan2024information,wan2024capacity}. Such keys can be efficiently generated through interactive key agreement among users, avoiding a dedicated key distribution server.
Moreover, SA under heterogeneous security constraints---referred to as weak security---was studied in~\cite{li2023weakly,li2025weakly,li2025collusionresilienthierarchicalsecureaggregation}. SA with user selection~\cite{zhao2022mds, zhao2023optimal} aims to recover the sum of inputs from a predefined subset of users, modeling partial participation in practical FL systems. Moreover, \hie\ \secagg\ (HSA) addresses SA in multi-layered networks, where users communicate with the  server via intermediate relays, and input security must be ensured against both the server and the relays.
SA was also studied in the context of secure gradient coding\cite{wan2022secure,zhou2025optimal,lu2024capacity,weng2025secure, weng2025cooperative}, where dataset redundancy is leveraged to achieve efficient communication and straggler tolerance.

It  can be seen that existing studies on  SA have mainly focused on centralized server-client architectures, sometimes with an additional relay layer. Motivated by the limited understanding of \itic SA in decentralized FL\cite{lalitha2018fully,he2018cola, beltran2023decentralized,rodio2025whisper}, this paper  studies the fundamental limits of \emph{\dsa} (DSA) in a $K$-user fully interconnected network. 
As illustrated in Fig.~\ref{fig:model}, each user holds a private input $W_k$ and communicates with all others over orthogonal, error-free broadcast channels. Each user aims to compute the global sum $\sum_{k=1}^KW_k$ while ensuring that no user, alone or colluding with up to $T$ others, learns any additional information about $(W_1,\ldots,W_K)$. To achieve   security, each user has a key $Z_k$ derived from a source key $\zsigma$ satisfying $H(Z_1,\ldots,Z_K|\zsigma)=0$, and transmits a message $X_k$ generated from $W_k$ and $Z_k$. Let $R_X$, $R_Z$, and $\rzsigma$ denote the per-user communication, individual and source key rates, respectively. Our goal is to characterize the optimal rate region $\Rc^*$, comprising all achievable rate triples $(R_X,R_Z,\rzsigma)$.

The major contribution of this work is a complete characterization of $\Rc^*$. Specifically, we show that
for the nontrivial scenario\footnote{For the scenario with $K=2$ or $T\ge K-2$, although the  \secty constraint (\ref{eq:security constraint}) can be satisfied  under  the proposed scheme, these cases are less interesting because  they degenerate to a $2$-user  \secagg problem. See a detailed explanation in Section \ref{subsec:infeasibility proof, converse}.} where there are at least  $K\ge 3$ users and at most $T\le K-3$ colluding users,
the optimal rate region is given by   $\Rc^*=\{(R_X, R_Z, \rzsigma):\rx\ge 1, \rz\ge 1, \rzsigma \ge K-1\}$.
That is, \emph{to securely compute one bit of the intended input sum (simultaneously at all users), each user must (i) transmit at least one \msg bit, (ii) hold at least one bit of secret key, and (iii) all users must collectively hold at least $K - 1$ independent key bits.}
The above result is achieved by an intuitive linear  \agg scheme, paired with a set of converse bounds that tightly establish the minimum \comm and key rates. Together, they provide the first \itic characterization of the fundamental limits of SA in fully decentralized settings.

\section{Problem Formulation}
\label{sec:problem formulation} 

\begin{figure}[t]
    \centering
    \includegraphics[width=0.44\textwidth]{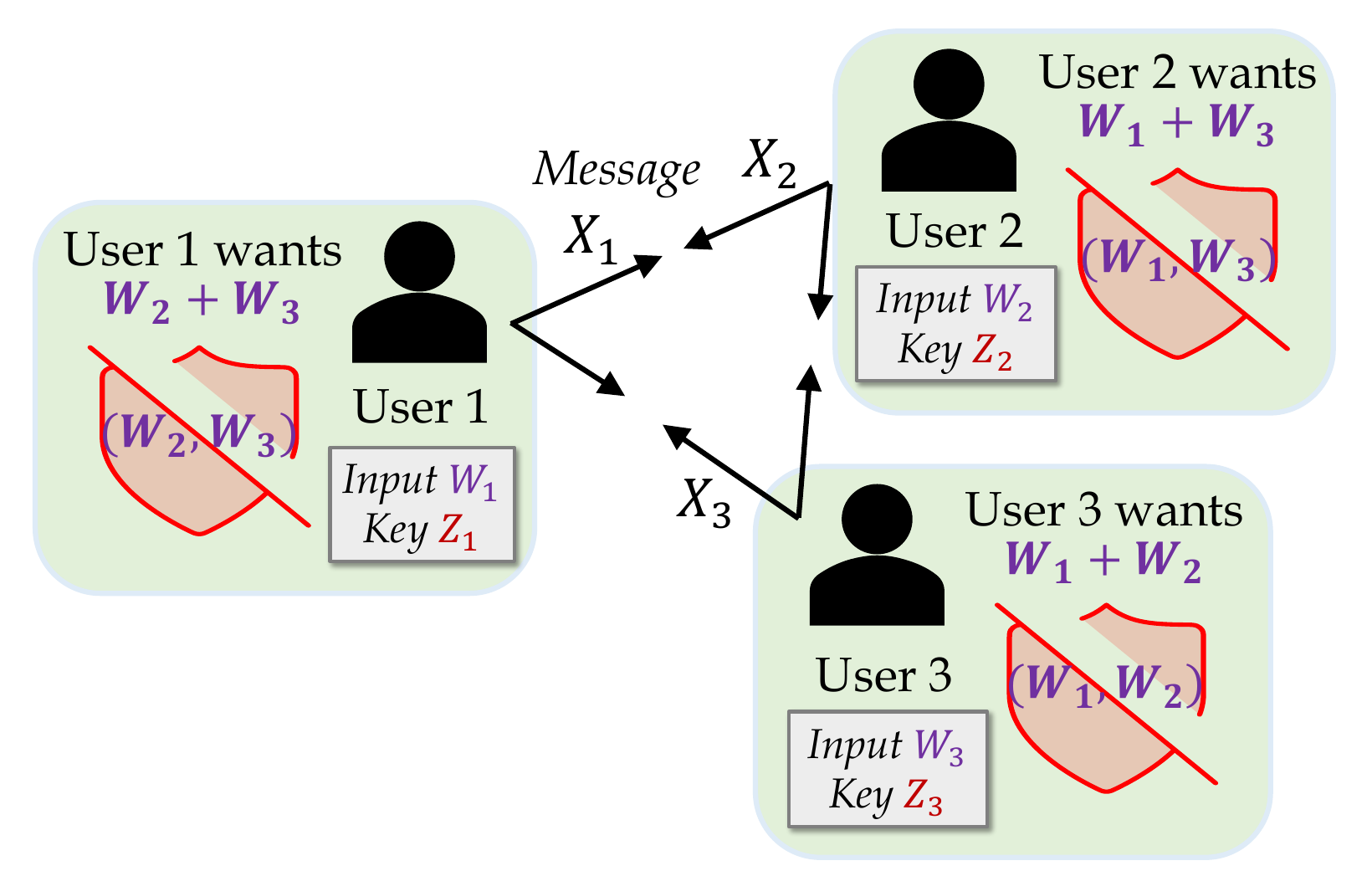}
    \vspace{-.2cm}
    \caption{\small DSA with  $K=3$ users. Each user wants to recover the sum of the inputs of the other two users.}
    \label{fig:model}
    \vspace{-.3cm}
\end{figure}

Consider a system of $K\ge 3$ users, where each user is connected to the remaining users through an error-free broadcast channel as shown in Fig.~\ref{fig:model}.
Each User $k \in [K]$ holds an \emph{input} $W_k$, consisting of $L$ \iid uniformly distributed bits\footnote{The \iid uniformity assumption on  the inputs 
is only necessary to establish the optimality of  our results. The proposed scheme works for arbitrarily distributed and \corr  inputs.}. This input serves as an abstraction of the locally trained ML models of the clients in federated learning (FL). To facilitate information-theoretic converse on the relevant \comm and secret key rates, we assume that the inputs of different users are mutually \indep, \ie, 
\begin{align}
\label{eq:input independence}
H(W_k) = L,\forall  k\in[K],\; H(W_{1:K}) =\sum_{k=1}^KH(W_k).
\end{align} 
Each user also possesses a secret key $Z_k$ (referred to as an \emph{\indiv key}) comprising $\lz$ bits.  The keys are assumed to be  \indep  of the inputs, namely,
\be 
\label{eq: indep. between keys and inputs}
H(Z_{1:K}, W_{1:K}) = H(Z_{1:K}) + \sum_{k=1}^KH(W_k).
\ee 
The \indiv keys $Z_{1:K}\eqdef \{Z_1, \cdots,Z_K\}$ can be  correlated in an arbitrary manner and are generated from a \emph{source key} variable $\zsigma$, which consists of $\lzsigma$ bits, such that
\be 
\label{eq: H(Z1,...,ZK|Zsigma)=0}
H\left(Z_{1:K}|\zsigma\right)=0.
\ee 
The source key $\zsigma$ represents the complete set of independent random key bits required to achieve security.
The individual keys are distributed by a trusted third-party entity in an offline fashion; specifically, each $Z_k$ is delivered to User $k$ privately prior to the aggregation process.

Each User $k\in[K]$ then generates a \msg  $X_k$ (with $L_X$ bits) based on $W_k$ and $Z_k$, and broadcasts the \msg to all remaining users. $X_k$ is assumed to be a deterministic function of $W_k$ and $Z_k$ so that 
\be
\label{eq:H(Xk|Wk,Zk)=0}
H(X_k|W_k,Z_k)=0,\; \forall  k\in[K]
\ee
We further assume that  the $K$ \msgs are transmitted over orthogonal channels (\eg, via TDMA, OFDM, etc.), so that  no interference occurs among them.
Upon  receiving the \msgs from all other users---and with access to its own input and secret key---each User $k$ should be able to recover the sum of all inputs, \ie, $\forall  k\in[K]$:
\begin{align}
\label{eq:recovery constraint}
& \mathrm{[Recovery]} \quad H\left(\inputsum \bigg|\{X_i\}_{i\in [K]\bksl \{ k\} }, W_k,Z_k\right  )=0,
\end{align}
Note that $X_k$ is not included in the  conditioning terms because it can be derived from $W_k$ and $Z_k$.
Moreover, since $W_k$ is available to User $k$, it only needs to recover $\sum_{i \in [K] \setminus \{k\}} W_i$.

\tbf{Security model.} 
We consider the \tit{honest-but-curious} \secty model where each user follows the prescribed \agg protocol faithfully (\ie, they do not tamper with their inputs or \msgs), but is curious to learn other users' inputs.
\Ip, each user must not learn anything about other users' inputs beyond  the aggregate sum (and whatever available through collusion) even if it can  collude with up to $T$ other users.
Here, \collu means that \user{k} may gain access to the inputs and keys of any subset of at most $T$ users $\Tc \subset [K]\bksl \{k\},|\Tc|\le T$. 
\Diff users may collude with \diff subsets of users. For ease of notation, let
$\Cc_{\Sc} \eqdef \{W_i,Z_i\}_{i\in \Sc}$
denote  the collection of inputs and keys of the users in $\Sc \subseteq [K]$. Using \muinfo,  the security constraint
can be expressed as
\begin{align}
\label{eq:security constraint}
& \mathrm{[Security]} \notag\\
& I\bigg( \{X_i\}_{i \in [K]\bksl \{k\}}; \{W_i\}_{i \in [K]\bksl \{k\}} \Big|
 \sum_{i=1}^K W_i, 
 W_k, Z_k, \Cc_\Tc   \bigg)=0,\notag\\
 & \hspace{3cm} \forall \Tc \subset [K]\bksl \{k\}, |\Tc|\le T, \forall k\in[K]
\end{align}
which imposes \indepce between  the received \msgs by  User $k$ and all other users' inputs  when conditioned on User $k$'s own input, secret key and the desired input sum. Note that  the above security constraint needs to be satisfied simultaneously at all users. To exclude the trivial $2$-user case, where each user must recover the other's input, we assume $K \ge 3$.

We study both the \comm and secret key generation aspects of  \dsa. \Msp, the  per-user 
\comm rate $ \rx  $  quantifies the communication efficiency  of  the \agg  process, and is defined as
\be
\label{eq: def comm rate}
R_X \eqdef {L_X}/{L},
\ee 
where $L_X$ denotes the number of bits contained in  $X_k$. 
The \indiv key rate $R_Z$ quantifies the number of \indep  secret key bits that each user must possess to ensure security,
\be
\label{eq: def indiv key rate}
R_Z \eqdef  {L_Z}/{L},
\ee 
where $L_Z$ denotes the number of bits contained in each $Z_k$.
The \skr  quantifies the total number of \indep key bits held collectively by all users, and is defined as 
\be
\label{eq: def source key rate} \rzsigma \eqdef  {\lzsigma}/{L},
\ee 
where $\lzsigma$ denotes the number of bits contained in the source key $\zsigma$. 
A rate triple $(\rx, \rz,\rzsigma)$ is said to be achievable if there exists a secure \agg scheme that \emph{simultaneously} achieves the \comm and key rates $\rx,\rz$, and $\rzsigma$, while satisfying both the recovery and the security constraints (\ref{eq:recovery constraint}) and (\ref{eq:security constraint}).
We aim to find the optimal rate region $\Rc^*$, defined as the closure of all achievable rate triples.

%%%%%%%%%%%% main result %%%%%%%%%%%%%%%
\section{Main Result}
\begin{theorem}
\label{thm:main thm}
For \dsa with $K\ge  3$ users and  a collusion threshold  when $ T \le  K- 3$, the optimal rate region   is given by
\be 
\label{eq:optimal rate region}
\Rc^* =\left\{ \left(\rx, \rz,\rzsigma\right) \left|  \begin{array}{c}
\rx \ge 1, \\
\rz \ge 1, \\
\rzsigma  \ge K-1
\end{array}
\right.   \right\}.
\ee
\vspace{-.2cm}
\end{theorem}

The \achvblty proof of \Thm \ref{thm:main thm} is presented in Section \ref{sec:proposed scheme}, where we propose a linear \agg scheme that achieves  the rates $\rx=1,\rz =1$, and $\rzsigma=K-1$. The converse proof is given  in Section \ref{sec: converse}, where we derive lower bounds $\rx\ge 1, \rz \ge 1$ and $\rzsigma\ge K-1$ using Shannon-theoretic proofs. Due to the  match of \achvblty and converse, the optimal region  can be established.

%%%%%%%%%%%%%%
\section{Proposed Scheme}
\label{sec:proposed scheme}
We first present an example to convey the intuition behind the proposed design, and then describe the general scheme.

\subsection{Motivating Examples}
\label{subsec:examples}

\begin{example}[$K=3,T=0$]
\label{example:K=3,T=0}
Consider $K=3$. Since $T\le  K-3=0$, no collusion can be  tolerated in this example. Let each input contain $L=1$ bit.  The source key is given by $\rzsigma=(N_1, N_2)$ where  $N_1$ and $N_2$ are two  \iid uniform random bits unknown to the server. Hence, the \skr is equal to $\rzsigma=2$. The \indiv keys are chosen as
\begin{align}
\label{eq:indiv keys,example 1}
Z_1 = N_1,\;
Z_2  = N_2, \;
Z_3 = -(N_1+N_2).
\end{align}
Since each key contains one bit, the \indiv key  rate is  $\rz=1$.
It can be seen that the keys have the zero-sum property, \ie,  $Z_1 + Z_2 + Z_3 = 0$, which is a necessary condition for recovery of the input sum.
The broadcast \msgs are chosen as
\begin{align}
\label{eq:msgs,example 1}
X_1 &= W_1+ Z_1 = W_1+N_1, \notag \\
X_2 &= W_2+Z_2=W_2+N_2, \notag \\
X_3 &= W_3 +Z_3=W_3-(N_1+N_2),
\end{align}
\ie, each \msg is a simple summation of  each user's input and key. Since each \msg contains $\lx =1$  bit,  the \comm rate is equal to $\rx=1$.

\tbf{Proof of security.} We now show that  the proposed scheme satisfies the security constraint (\ref{eq:security constraint}). \Wlog, let us consider User $1$. An intuitive proof is provided \af: 
Suppose User $1$ recovers the desired input sum
$W_2+W_3$ by linearly combining $X_2,X_3$ and $Z_1$ with coefficients $\ell_1, \ell_2$ and $\ell_3$, \resp. That is, 
$ \ell_1X_2+\ell_2X_3 + \ell_3Z_1
=  \ell_1W_2+\ell_2W_3 +  (\ell_3-\ell_2)N_1 +  (\ell_1-\ell_2)N_2$.
\if0
\begin{align}
&  \ell_1X_2+\ell_2X_3 + \ell_3Z_1\notag\\
& =  \ell_1W_2+\ell_2W_3 +  (\ell_3-\ell_2)N_1 +  (\ell_1-\ell_2)N_2.
\end{align}
\fi 
Because $N_1$ and $N_2$ are two  \indep bits, to cancel out  the key component $(\ell_3-\ell_2)N_1 +  (\ell_1-\ell_2)N_2$, the coefficients must be zero, \ie,
$ \ell_3-\ell_2=\ell_1-\ell_2=0$, suggesting $\ell_1=\ell_2=\ell_3$.
Therefore, any key-canceling linear combination\footnote{A key-canceling linear combination ensures that all key variables cancel out completely, leaving only the input component intact.} $\bm{\ell}\eqdef (\ell_1, \ell_2, \ell_3)$ must take the form $\bm{\ell}=(\ell, \ell, \ell), \ell \in \{0,1\}$. As a  result,
\be
\label{eq:l*X=l(W2+W3),example}
\bm{\ell}(X_2,X_3,Z_1)^{\rm  T}=\ell(W_2+W_3),
\ee 
which is a function of the desired input sum $W_2+W_3$. (\ref{eq:l*X=l(W2+W3),example})  implies that any key-canceling linear combination of the received \msgs and User $1$'s key reveals no additional information about  $(W_2, W_3)$ beyond their sum $W_2+W_3$, thereby ensuring security. 
A rigorous proof of security through zero mutual \info is given in Section \ref{subsec:general scheme}.
\hfill $\lozenge$
\end{example}

\subsection{General Scheme}
\label{subsec:general scheme}
Suppose each input has $L=1$ bit. The source key is 
\be
\label{eq: source key, gen scheme}
\zsigma =\left(N_1, \cdots, N_{K-1}\right),
\ee 
where $N_1, \cdots, N_{K-1}$ are $K-1$ \iid uniform random bits. Therefore,  $\rzsigma=H(N_1, \cdots, N_{K-1})/L=K-1$. The \indiv keys are chosen as 
\begin{align}
\label{eq: key design, gen scheme}
Z_k & = N_k, \; k \in[K-1] \notag  \\
Z_K & = -(N_1+\cdots+ N_{K-1}).
\end{align}
This key design has two properties: first, the \indiv keys have a zero sum, \ie, $Z_1+\cdots+Z_{K}=0$; second, any subset of no more than $K-1$ keys are mutually \indep. The first property is necessary for  the input sum recovery by the server, while the second property is necessary for security. With the above key design, User $k\in [K]$ broadcasts
\be
\label{eq:msg design, gen scheme}
X_k = W_k + Z_k
\ee 
to all other users.

\subsubsection{Input Sum Recovery}
Each user adds up all received \msgs and plugs in its own input and secret key to decode  the desired input sum.  \Ip,  User $k$ performs 
\begin{align}
% \label{eq:input sum recovery,gen scheme}
 & W_k +Z_k + \sum_{i\in [K]\bksl \{k\}  }X_i      = \sum_{k=1}^KW_k + \sum_{k=1}^K Z_k \overset{(\ref{eq: key design, gen scheme})}{=} \sum_{k=1}^KW_k\notag
\end{align}
to recover the  sum of inputs of all users.

\subsubsection{Performance} The \achved rates are $\rx=1, \rz=1$ and $\rzsigma=K-1$.

\subsubsection{Proof of Security}
\label{subsubsec:proof of security}
With  the proposed secret key and \msg design, we prove that the security constraint in (\ref{eq:security constraint}) can be simultaneously satisfied for all users. Specifically,  consider User $k$ and any  colluding user set $\Tc \subset [K]\bksl\{k\}$ where $|\Tc|\le K-3$. \Feon, let us denote $N_K \eqdef -(N_1+\cdots+N_{K-1})$ so that  $Z_k=N_k, \forall k\in[K]$, and $\overline{\Tc}\eqdef ([K]\bksl \{k\})\bksl \Tc$. 
We have
{\setlength{\jot}{.5pt}
\begin{subequations}
\label{eq:proof of security, general scheme}
\begin{align}
& I\left( \{X_i\}_{i \in [K]\bksl \{k\}}; \{W_i\}_{i \in [K]\bksl \{k\}} \bigg|
 \sum_{i=1}^K W_i, 
 W_k, Z_k, \Cc_\Tc   \right)\notag \\
 &\overset{(\ref{eq: key design, gen scheme})}{=}    I\left( \{W_i+N_i\}_{i \in \overline{\Tc}   }; \{W_i\}_{i\in \overline{\Tc}} \bigg|
 \sum_{i \in  \overline{\Tc}  } W_i, 
 \Cc_{\Tc \cup  \{k\}   }   \right)\label{eq:step0,proof of security, general scheme} \\
 & 
= H\left( \{W_i+N_i\}_{i \in \overline{\Tc}   } \bigg|
 \sum_{i \in  \overline{\Tc}  } W_i, 
 \Cc_{\Tc \cup  \{k\}   }   \right)\notag\\
& \;\;- H\left( \{W_i+N_i\}_{i \in \overline{\Tc}   } \bigg|
 \sum_{i \in  \overline{\Tc}  } W_i, 
 \Cc_{\Tc \cup  \{k\}   }, \{W_i\}_{i\in \overline{\Tc}}   \right)\\
& 
\overset{(\ref{eq: key design, gen scheme})}{=}  H\left( \{W_i+N_i\}_{i \in \overline{\Tc}   } \bigg|
 \sum_{i \in  \overline{\Tc}  } W_i, \sum_{i \in \overline{\Tc}}N_i,  \Cc_{\Tc \cup  \{k\}   }   \right)\notag\\
& \quad 
- H\left( \{N_i\}_{i \in \overline{\Tc}   } \bigg|
 \sum_{i \in  \overline{\Tc}  } W_i, 
 \Cc_{\Tc \cup  \{k\}   }, \{W_i\}_{i\in \overline{\Tc}}   \right)\label{eq:step1,proof of security, general scheme}\\
& \overset{(\ref{eq: indep. between keys and inputs})}{\le }  
H\left( \{W_i+N_i\}_{i \in \overline{\Tc}   } \bigg|
 \sum_{i \in  \overline{\Tc}  } (W_i+N_i)  \right)\notag\\
& \quad 
- H\left( \{N_i\}_{i \in \overline{\Tc}   } \big|
 \{N_i\}_{i\in \Tc \cup  \{k\}   }   \right)\label{eq:step2,proof of security, general scheme}\\
& =  
H\left( \{W_i+N_i\}_{i \in \overline{\Tc}   } \right) 
-
H\left(  \sum_{i \in  \overline{\Tc}  } (W_i+N_i)\right)
\notag\\
& \quad 
- 
H\left( \{N_i\}_{i \in [K]   }   \right)
+H\left(
 \{N_i\}_{i\in \Tc \cup  \{k\}   }   \right)
 \label{eq:step3,proof of security, general scheme}\\
 & 
 \overset{(\ref{eq:input independence}), (\ref{eq: indep. between keys and inputs}),(\ref{eq: key design, gen scheme})}{=}
|\overline{\Tc}|L-L-(K-1)L+|\Tc \cup \{k\}|L \label{eq:step4,proof of security, general scheme}\\
&=
\left(|\overline{\Tc}|+|\Tc \cup \{k\}|-K     \right)L=0,
\end{align}
\end{subequations}}
where in (\ref{eq:step0,proof of security, general scheme}), we substitute the message and key constructions from  (\ref{eq:msg design, gen scheme}) and (\ref{eq: key design, gen scheme}), \resp, and exclude the conditioning inputs and keys $ \Cc_{\Tc \cup  \{k\}   }= \{W_i,N_i\}_{i\in \Tc\cup  \{k\} }$ from the relevant terms. (\ref{eq:step1,proof of security, general scheme}) is due to the zero-sum property of the keys, \ie,
$\sum_{i \in \overline{\Tc}}N_i= - \sum_{i\in \Tc \cup  \{k\}} N_i  $, which can be derived from $\Cc_{\Tc\cup \{k\}}$. (\ref{eq:step2,proof of security, general scheme}) is due to the \indepce of the inputs and keys (See (\ref{eq: indep. between keys and inputs})). 
(\ref{eq:step3,proof of security, general scheme})
is because 
$H\big(  \sum_{i \in  \overline{\Tc}  } (W_i+N_i)| \{W_i+N_i\}_{i \in \overline{\Tc}   }\big)=0$, and also  the fact that  $ \Tc \cup \overline{\Tc}\cup  \{k\}=[K]$. (\ref{eq:step4,proof of security, general scheme})  follows from the uniformity of both the inputs and keys, as well as their mutual independence. Since \muinfo is non-negative, we have $I\big( \{X_i\}_{i \in [K]\bksl \{k\}}; \{W_i\}_{i \in [K]\bksl \{k\}}|
 \sum_{i=1}^K W_i, 
W_k, Z_k, \Cc_\Tc \big)=0$. Since this condition holds simultaneously for all users, security is proved.

%%%%%%%%%%%%%%%%%%%%%%%
\section{Converse}
\label{sec: converse}
This section presents an entropic converse proof to show that for any DSA scheme, the rates must be lower bounded by $\rx  \ge 1, \rz \ge 1$ and $\rzsigma \ge K-1$, therefore establishing the optimality of the proposed scheme  in Section \ref{sec:proposed scheme}.

\subsection{Trivial Regime: $K=2$ or $T\ge K-2$}
\label{subsec:infeasibility proof, converse}
When there are only two users, computing the sum $W_1 + W_2$ effectively requires each user to recover the other user's input, leaving nothing to be concealed. Hence, no meaningful security can be provided. Similarly, if a user can collude with $T=K-2$  other users, it will learn the inputs and keys of all users except one. \Aar, the problem reduces to the $2$-user \agg scenario,  in which no meaningful security can be achieved. \Iwf, we assume $K\ge 3$ and $T\le K-3 $.

\subsection{Lemmas and Corollaries}
\label{subsec:lemmas&corollaries}
This section presents several lemmas and corollaries that form the building blocks of the converse proof. Detailed proofs are omitted due to space constraints and can be found in the extended version of this paper \cite{zhang2025information}.

The first lemma states that each message $X_k$ contains at least $L$ independent bits, even when the inputs and keys of all other users are known.

\begin{lemma}
\label{lemma: H(Xk|(Wi,Zi) all other i)}
\emph{For any $k\in [K]$,  it holds that
\be 
\label{eq: H(Xk|(Wi,Zi) all other i), lemma}
H\left(X_k| \{W_i, Z_i\}_{i\in [K]\bksl \{k\}  } \right) \ge H(W_k)=L.   
\ee 
}
\end{lemma}

Let $\Tc_{(k)} \subset [K]\bksl \{k\}$, with $|\Tc_{(k)} |\le K-3$, denote a subset of users that may collude with ser $k$. We define its complement---relative to  the full set  $[K]\bksl\{k\}$---as $ \overline{\Tc}_{(k)}  \eqdef \left([K]\bksl\{k\}\right)\bksl \Tc_{(k)} $.
A direct implication of Lemma~\ref{lemma: H(Xk|(Wi,Zi) all other i)} is the following corollary, which states that the joint entropy of the \msgs $\{X_i\}_{i\in \overline{\Tc}_{(k)} }$, when conditioned on the inputs and keys of the colluding set $\Tc_{(k)}$ and User $k$, is at least $|\overline{\Tc}_{(k)}|L$.

\begin{corollary}
\label{corollary1}
\emph{For any colluding user set $\Tc_{(k)} \subset [K]\bksl \{k\}$ and its complement $\overline{\Tc}_{(k)}$,  it holds that}
\be
\label{eq:corollary1}
H\left( \{X_i\}_{i\in \overline{\Tc}_{(k)} } \big| \{W_i,Z_i\}_{i\in\Tc_{(k)}}, W_k,Z_k       \right) \ge |\overline{\Tc}_{(k)} |L.
\ee 
\end{corollary}

The following lemma states that any \msg  $X_k$ must be \indep of the \corrspdg input $W_k$, when conditioned on the input $W_{k^\prime}$ and key $Z_{k^\prime}$ of any other User $k^\prime \ne k$. This follows from the security constraint imposed on User $k^\prime$, which  requires that $W_k$ to be fully protected by $Z_k$. In other words, $X_k$ must be \indep of $W_k$,
which translates to $I(X_k; W_k|W_{k^\prime}, Z_{k^\prime})=0,\forall k\ne k^\prime$. The conditioning on $W_{k^\prime}$ and $Z_{k^\prime}$
reflects the fact that User $k^\prime$ has access to  this \info, and hence it must be treated as known when evaluating the mutual information.

\begin{lemma}
\label{lemma: I(Xk;Wk|Wk',Zk')=0}
\emph{For any pair of users $(k, k^\prime)$, it holds that
\begin{align}
\label{eq:lemma: I(Xk;Wk|Wk',Zk')=0}
I\left(X_k; W_k |W_{k^\prime}, Z_{k^\prime}     \right)=0.
\end{align}
}
\end{lemma}
\

\begin{lemma}
\label{lemma: I((Xi)_Tkc;(Wi)_Tkc|Wk,Zk,(Wi,Zi)_Tk)=L}
\emph{
For any $k\in [K]$, any colluding user set $\Tc_{(k)} \subset [K]\bksl \{k\}$ and its complement $\overline{\Tc}_{(k)}$, the following equality holds:}
\begin{align}
\label{eq:I((Xi)_Tkc;(Wi)_Tkc|Wk,Zk,(Wi,Zi)_Tk)=L, lemma}
 & I\left(\{X_i\}_{i\in \overline{\Tc}_{(k)}}; 
\{W_i\}_{i\in \overline{\Tc}_{(k)}} \big|\{W_i,Z_i\}_{i\in \Tc_{(k)}}, W_k,Z_k\right)=L.
\end{align} 
\end{lemma}

Lemma~\ref{lemma: I((Xi)_Tkc;(Wi)_Tkc|Wk,Zk,(Wi,Zi)_Tk)=L} implies that, from  the perspective of any User $k$, even if it colludes with  a  subset of users $\Tc_{(k)}\subset[K]\bksl \{k\}$, the only \info that can be  inferred about the inputs $\{W_i\}_{i\in \overline{\Tc}_{(k)}} $ is their sum. This reflects the core security requirement imposed on User $k$: it must not learn anything beyond the sum of inputs and the information already available through collusion, regardless of which subset of users it colludes with.

We also have the following lemma, which asserts  that the joint entropy of the  \indiv  keys in any $\overline{\Tc}_{(k)}$, when conditioned on all other keys in $\Tc_{(k)}\cup \{k\}$, is at least $(|\overline{\Tc}_{(k)}|-1)L$ for any User $k\in [K]$.
\begin{lemma}
\label{lemma:H((Zi)_Tkc|(Zi,Wi)_Tk, Wk,Zk)>=(K-|Tkc|-2)L}
\emph{
For any set of colluding users $\Tc_{(k)}\subseteq [K]\bksl\{k\}$ and its complement $\overline{\Tc}_{(k)}=( [K]\bksl\{k\})\bksl \Tc_{(k)}$, it holds that}
\be
\label{eq:H((Zi)_Tkc|(Zi,Wi)_Tk, Wk,Zk)>=(K-|Tkc|-2)L, lemma}
H\left( \{Z_i\}_{ i\in\overline{\Tc}_{(k)}  } \big| \{Z_i\}_{i\in \Tc_{(k)}},Z_k   \right)
\ge \left(K-2-|\Tc_{(k)}|\right)L.
\ee
\end{lemma}

\begin{figure}[ht]
    \centering
    \includegraphics[width=0.36\textwidth]{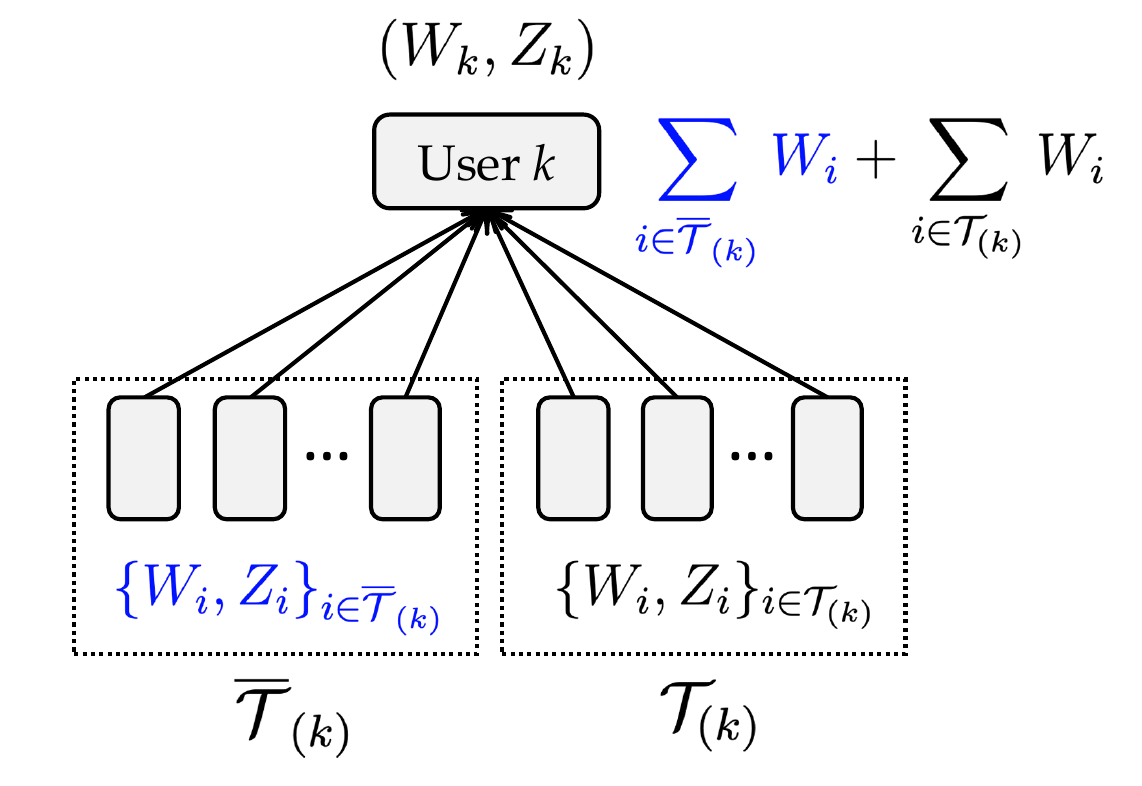}
    \vspace{-.4cm}
    \caption{\Agg at User $k$ with colluding user set $\Tc_{(k)}$. Because the inputs $\{W_i\}_{ i\in \Tc_{(k)}}$ are already known via collusion, User $k$ only needs to recover the sum of the remaining inputs $\sum_{i\in \overline{\Tc}_{(k)}  }W_i$.}
    \label{fig:fig2}
    \vspace{-.2cm}
\end{figure}

An intuitive interpretation of Lemma~\ref{lemma:H((Zi)_Tkc|(Zi,Wi)_Tk, Wk,Zk)>=(K-|Tkc|-2)L} is \af.
Let us consider the \agg at User $k$ as shown in Fig.~\ref{fig:fig2}.
Since the inputs and keys $ \{W_i,Z_i\}_{i\in \Tc_{(k)}}$ are already known to User $k$ through collusion, it only needs  to recover the sum of inputs $\sum_{i\in \overline{\Tc}_{(k)}}W_i$ of the remaining users $\overline{\Tc}_{(k)}$. To ensure the security of these inputs against User $k$, a fundamental result by Zhao and Sun~\cite{zhao2023secure} states that the users in $\overline{\Tc}_{(k)}$ must collectively hold at least $(|\overline{\Tc}_{(k)}|-1)L$ \indep key bits. Provided that User $k$  also possesses  $Z_k$ in addition to the colluded keys $\{Z_i\}_{i\in \Tc_{(k)}}$, we have 
$
H( \{Z_i\}_{ i\in\overline{\Tc}_{(k)}  } | \{Z_i\}_{i\in \Tc_{(k)}},Z_k )
\ge (|\overline{\Tc}_{(k)}|-1)L=(K-2-|\Tc_{(k)}|)L
$.

\subsection{Lower Bounds on \Comm and Key Rates}
\label{subsec:lower bounds on comm and key rates}
Equipped with the lemmas in the previous section, we now derive the lower bounds on the \comm and key rates. Since these bounds match the \achvb rates in Section~\ref{sec:proposed scheme}, the optimality of the proposed scheme can be established.

\subsubsection{Proof of \Comm Rate $\rx \ge 1$}
\label{subsubsec:proof of Rx>=1}
Consider any User $k\in[K]$. 
A straightforward application of Lemma~\ref{lemma: H(Xk|(Wi,Zi) all other i)} gives
\begin{subequations}
\begin{align}
L_X \ge H(X_k) & \ge H\left(X_k |\{W_i, Z_i\}_{i\in [K]\bksl \{k\}   } \right)\overset{(\ref{eq: H(Xk|(Wi,Zi) all other i), lemma})}{\ge } L\\
  \Rightarrow \rx  & \eqdef {L_X}/{L} \ge  1.
\end{align}
\end{subequations}

\subsubsection{Proof of \Indiv Key Rate $\rz \ge 1$}
\label{subsubsec:proof of Rz>=1}
Consider any pair of users $(k,k^\prime)$. We have
{\setlength{\jot}{.5pt}
\begin{subequations}
\label{eq:proof of Rz>=1,converse}
\begin{align}
L_Z & \ge H(Z_k) \ge   H\left(Z_k|W_k, W_{k^\prime}, Z_{k^\prime} \right  )\\
 & \ge I\left(Z_k; X_k|W_k, W_{k^\prime}, Z_{k^\prime}   \right)\\
 & = H\left(X_k|W_k, W_{k^\prime}, Z_{k^\prime} \right) - \underbrace{ H\left(X_k|W_k, Z_k,W_{k^\prime}, Z_{k^\prime} \right)}_{\overset{(\ref{eq:H(Xk|Wk,Zk)=0})}{=} 0   }\notag\\
& = H\left(X_k|W_{k^\prime}, Z_{k^\prime} \right) - \underbrace{ I\left(X_k;W_k|W_{k^\prime}, Z_{k^\prime} \right)}_{\overset{(\ref{lemma: I(Xk;Wk|Wk',Zk')=0})}{=}0 } \label{eq:step0,proof of Rz>=1,converse}\\
& \ge  H\left(X_k|\{W_i,Z_i\}_{ i\in  [K]\bksl\{k\}  } \right)\label{eq:step1,proof of Rz>=1,converse}\\
& \overset{(\ref{eq: H(Xk|(Wi,Zi) all other i), lemma})}{\ge}L
\Rightarrow \rz  \eqdef {L_Z}/{L}\ge 1,
\label{eq:step2,proof of Rz>=1,converse}
% \label{eq:step3,proof of Rz>=1,converse}
\end{align}
\end{subequations}}
where Lemma~\ref{lemma: I(Xk;Wk|Wk',Zk')=0} is applied in (\ref{eq:step0,proof of Rz>=1,converse}). (\ref{eq:step1,proof of Rz>=1,converse}) is because $k^\prime \in[K]\bksl\{k\} $ since $k\ne k^\prime$. Moreover, Lemma~\ref{lemma: H(Xk|(Wi,Zi) all other i)} is applied in (\ref{eq:step2,proof of Rz>=1,converse}).

\subsubsection{Proof of Source Key Rate $\rzsigma \ge K-1$}
\label{subsubsec:proof of Rzsigma>=1} 
Consider any $k\in[K]$. By setting the colluding user set as $\Tc_{(k)}= \emptyset$ (so that $\overline{\Tc}_{(k)}=[K]\bksl\{k\}$) in Lemma~\ref{lemma:H((Zi)_Tkc|(Zi,Wi)_Tk, Wk,Zk)>=(K-|Tkc|-2)L}, we obtain
\be
\label{eq:eq1,proof of Rzsigma>=K-1}
H\left( \{Z_i\}_{i\in [K]\bksl\{k\}} |Z_k   \right) \ge(K-2)L.
\ee 
Then,
{\setlength{\jot}{.5pt}
\begin{subequations}
\label{eq:proof of Rzsigma>=K-1}
\begin{align}
\lzsigma \ge H\left(\zsigma  \right)
 &  \overset{(\ref{eq: H(Z1,...,ZK|Zsigma)=0})}{=}H\left(\zsigma  \right) + H\left(Z_{1:K}|\zsigma  \right)\label{eq:step0,proof of Rzsigma>=K-1} \\
 & = H\left(Z_{1:K}, \zsigma\right)\\
 & \ge H(Z_{1:K})\\
 & = H(Z_k) + H\left( \{Z_i\}_{i\in [K]\bksl\{k\}} |Z_k   \right)\\
& \overset{(\ref{eq:proof of Rz>=1,converse}),(\ref{eq:eq1,proof of Rzsigma>=K-1})}{\ge }
L + (K-2)L\\
&= (K-1)L\\
\Rightarrow & \rzsigma  \eqdef {\lzsigma}/{L} \ge K-1,
\end{align}
\end{subequations}}
where (\ref{eq:step0,proof of Rzsigma>=K-1}) is because the \indiv keys are generated from the source key (see (\ref{eq: H(Z1,...,ZK|Zsigma)=0})). \Aar, we proved  $\rzsigma \ge K-1$.

\section{Conclusion}
In this paper, we studied the problem of collusion-resilient \dsa (DSA) from an \itic perspective. Specifically, users in a fully connected network collaboratively compute the sum of all private inputs under a security constraint that prevents any additional information leakage, even under collusion.
We characterized the optimal rate region that specifies the minimum \simuly \achvb \comm and  secret key rates. Future work may extend \secagg to more general decentralized topologies with potentially unreliable communication channels.

\section*{Acknowledgement}

The work of X. Zhang and G. Caire was supported by the Gottfried Wilhelm Leibniz-Preis 2021 of the German Science Foundation (DFG). 
The work of Z. Li was supported in part by the National Natural Science Foundation of China (Grant No. 62401266), and the Guangxi Natural Science Foundation (Grant No. 2025GXNSFBA069315), and the Jiangsu Natural Science Foundation (Grant No. BK20241452).
The work of K. Wan was supported in part by the NSFC under Grant 62571206 and  Wuhan Chen Guang Program under Grant 2024040801020211.

% \clearpage % Or \newpage
\bibliographystyle{IEEEtran}
\bibliography{references_secagg.bib}

\end{document}